\documentclass{elsart}

\usepackage{graphicx}

\usepackage{hyperref}
\usepackage{amssymb}

\begin{document}

\DeclareGraphicsExtensions{.eps, .jpg}

\def\hslash{\hbar}
\def\imag{i}
\def\grad{\vec{\nabla}}
\def\div{\vec{\nabla}\cdot}
\def\curl{\vec{\nabla}\times}
\def\DDt{\frac{d}{dt}}
\def\ddt{\frac{\partial}{\partial t}}
\def\ddx{\frac{\partial}{\partial x}}
\def\ddy{\frac{\partial}{\partial y}}
\def\lap{\nabla^{2}}
\def\divv{\vec{\nabla}\cdot\vec{v}}
\def\gradS{\vec{\nabla}S}
\def\vvec{\vec{v}}
\def\wc{\omega_{c}}
\def\<{\langle}
\def\>{\rangle}
\def\Tr{{\rm Tr}}
\def\Csch{{\rm csch}}
\def\Coth{{\rm coth}}
\def\Tanh{{\rm tanh}}
\def\g2{g^{(2)}}

\begin{frontmatter}

\title{Frenkel Exciton Model of Ultrafast Excited State Dynamics in AT DNA Double Helices}
\author{Eric R. Bittner}
\address{Department of Chemistry and the Texas Center for Superconductivity, 
University of Houston, Houston TX 77204}
\ead{bittner@uh.edu}

\begin{abstract}
Recent ultrafast experiments have implicated intrachain base-stacking
rather than base-pairing as the crucial factor in determining the fate
and transport of photoexcited species in DNA chains. An important
issue that has emerged concerns whether or not a Frenkel excitons is
sufficient one needs charge-transfer states to fully account for the
dynamics.  In we present an $SU(2)\otimes SU(2)$ lattice model which
incorporates both intrachain and interchain electronic interactions to
study the quantum mechanical evolution of an initial excitonic state
placed on either the adenosine or thymidine side of a model B DNA
poly(dA).poly(dT) duplex.  Our calculations indicate that over several
hundred femtoseconds, the adenosine exciton remains a cohesive
excitonic wave packet on the adenosine side of the chain where as the
thymidine exciton rapidly decomposes into mobile electron/hole pairs
along the thymidine side of the chain.  In both cases, the very little
transfer to the other chain is seen over the time-scale of our
calculations.  We attribute the difference in these dynamics to the
roughly 4:1 ratio of hole vs. electron mobility along the thymidine
chain.  We also show that this difference is is robust even when 
structural fluctuations are introduced in the form of static 
off-diagonal disorder. 
\end{abstract}

\end{frontmatter}

\section{Introduction}
\label{intro}
For all life-forms on Earth with the exception of certain viruses,
genetic information is carried within the cellular nucleus via strands
of strands of deoxyribonucleic acid (DNA).  The genetic information
itself is encoded in the specific sequence of the nucleic acid bases:
adenine (A), thymine (T), guanine (G), and cytidine (C).  DNA is also
a strong absorber of ultraviolet light leaving it highly susceptible
to photomutagenic damage with the primary photoproducts being
bipyrimidine dimers linking neighboring T bases.  For all organisms,
this susceptibility is compensated for in part through enzymatic
repair actions that remove damaged segments along one strand using the
complementary strand as a template for replacement.  Such repair
mechanisms are quite costly energetically.  Remarkably, however, DNA
is intrinsically photochemically stable as single bases are able to
rapidly convert photoexcitation energy into thermal energy on a
picosecond time scale through non-radiative electronic processes.
What remains poorly understood, is the role that base-pairing and
base-stacking plays in the transport and migration of the initial
excitation along the double helix.  Clearly, such factors are
important since the UV absorption of DNA largely represents the
weighted sum of the absorption spectra of it constituent bases whereas
the distribution of lesions formed as the result of photoexcitation
are generally not uniformly distributed along the chain itself and
depend strongly upon sequence, suggesting some degree of coupling
between bases.\cite{Markovitsi:2005}

Given the importance of DNA in biological system and its emerging role
as a scaffold and conduit for electronic transport in molecular
electronic devices, \cite{Kelley:1999} DNA in its many forms is a well
studied and well characterized system.  What remains poorly
understood, however, is the role that base-pairing and base-stacking
plays in the transport and migration of the initial excitation along
the double helix.\cite{Crespo-Hernandez:2005,Markovitsi:2005} Such
factors are important since the UV absorption of DNA largely
represents the weighted sum of the absorption spectra of it
constituent bases whereas the distribution of lesions formed as the
result of photoexcitation are generally not uniformly distributed
along the chain itself and depend strongly upon sequence, suggesting
some degree of coupling between bases.\cite{Markovitsi:2005}

Recent work by various groups has underscored the different roles that
base-stacking and base-pairing play in mediating the fate of an
electronic excitation in DNA.
\cite{Markovitsi:2005,Crespo-Hernandez:2005} Over 40 years ago,
L{\"o}wdin discussed proton tunneling between bases as a excited state
deactivation mechanism in DNA\cite{Lowin:1963} and evidence of this
was recently reported by Schultz {\em et al.} \cite{Schultz:2004} In
contrast, however,ultrafast fluorescence of double helix
poly(dA).poly(dT) oligomers by Crespo-Hernandez et
al.\cite{Crespo-Hernandez:2005} and by Markovitsi {\em et al.}
\cite{Markovitsi:2005} give compelling evidence that base-stacking
rather than base-pairing largely determines the fate of an excited
state in DNA chains composed of A and T bases with long-lived
intrastrand states forming when ever A is stacked with itself or with
T.  However, there is considerable debate regarding whether or not the
dynamics can be explained via purely Frenkel exciton models
~\cite{Emanuele:2005a,Emanuele:2005,Markovitsi:2006} or whether
charge-transfer states play an intermediate
role. \cite{Crespo-Hernandez:2006}

\begin{figure}[t]
\begin{center}
\includegraphics[width=0.65\columnwidth]{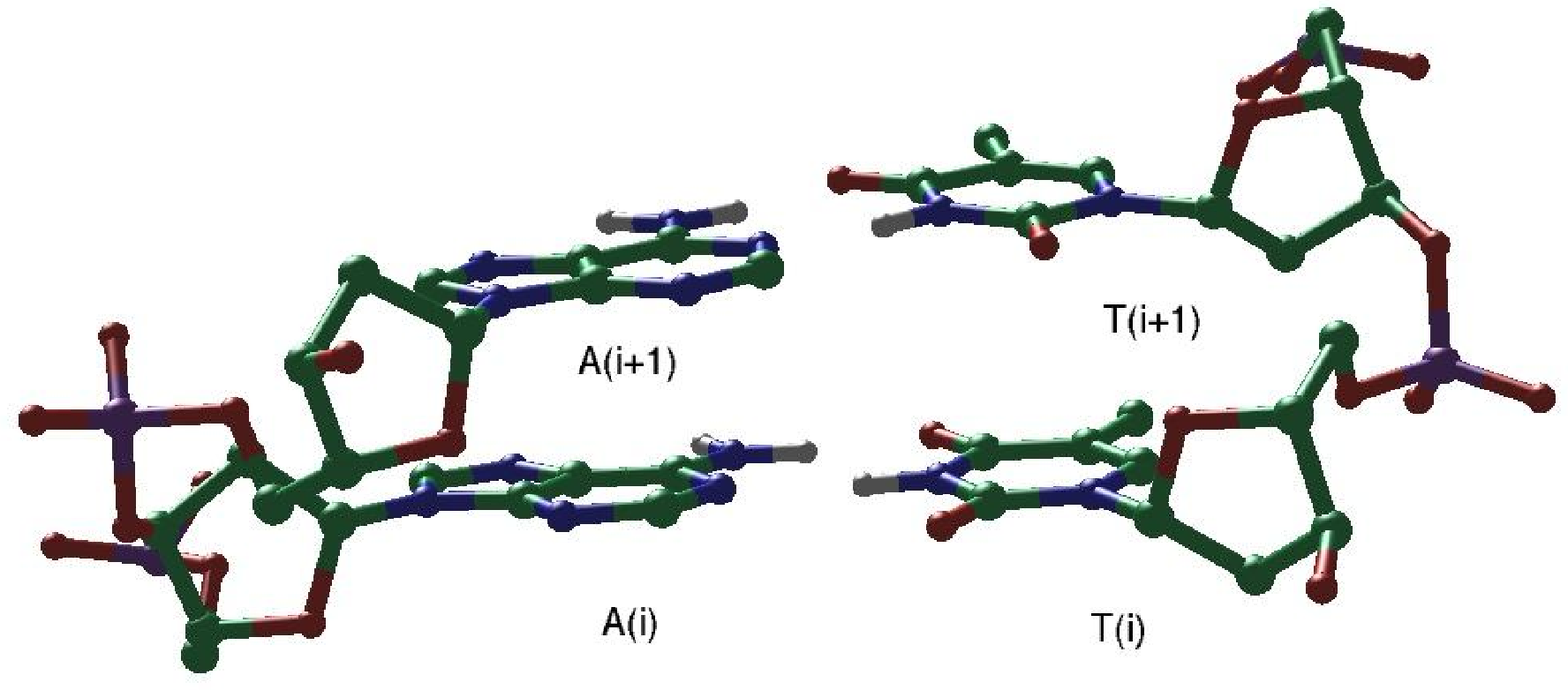}
\includegraphics[height=2.5in]{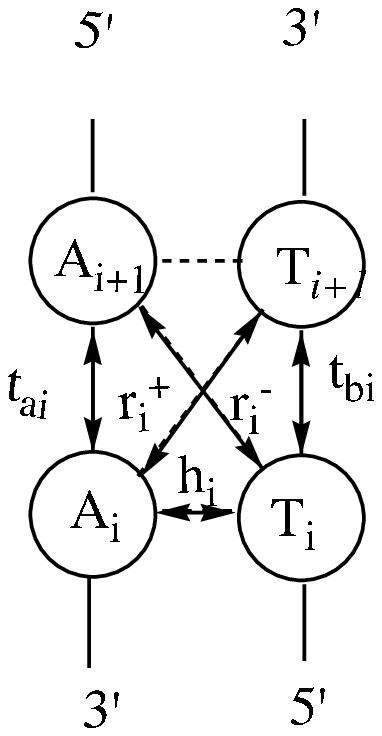}
\end{center}
\caption{Three dimensional structure of stacked A-T base pairs along with the 
corresponding lattice model. }
\label{fig1}
\end{figure}

Here we report on a series of quantum dynamical calculations that
explore the fate of a localized exciton placed on either the A side or
T side of the B DNA duplex poly(dA)$_{10}$.poly(dT)$_{10}$.  Our
theoretical model is based upon a $SU(2)\otimes SU(2)$ lattice model
we recently introduced~\cite{Bittner:094909} that consists of
localized hopping interactions for electrons and holes between
adjacent base pairs along each strand ($t_{aj}$) as well as
cross-strand terms linking paired bases ($h_i$) and ``diagonal'' terms
which account for the $\pi$ stacking interaction between base $j$ on
one chain and base $j\pm 1$ on the other chain ($r^\pm_i$) in which
$r^-_j$ denotes coupling in the 5'-5' direction and $r^+_j$ coupling
in the 3'-3' direction.  Fig. ~\ref{fig1} shows the three-dimensional
structure of poly(dA)$_{10}$.poly(dT)$_{10}$ and the topology of the
equivalent lattice model.    We also consider here the role of geometric
or structural fluctuations in the electronic dynamics. 

\section{Theoretical Model}

Taking each link as Fig. 1  as a
specific electron, hole, or excitonic, hopping term, we arrive at the following
single particle Hamiltonian,
\begin{eqnarray}
h_{1} = \sum_j \epsilon_j \hat{\psi}_j^\dagger\hat\psi_j
+ t_j (\hat\psi_{j+1}^\dagger\hat\psi_j + \hat\psi_{j}^\dagger\hat\psi_{j+1}) )
+ h_j\overline{\psi}_j\hat\psi_j \nonumber \\
+\hat\psi_{j+1}^\dagger(r_j^+\hat\gamma_+ + r_j^-\hat\gamma_-)\hat\psi_j
+\hat\psi_{j}^\dagger(r_j^+\hat\gamma_+ + r_j^-\hat\gamma_-)\hat\psi_{j+1},
\label{bandop}
\end{eqnarray}
where $\hat\psi_j^\dagger$ and $\hat\psi_j$ are $SU(2)$ spinors that
act on the ground-state to create and remove an electron (or hole) on
the $j$th adenosine or thymidine base along the chain.  The
$\hat\gamma$ operators are the $2\times 2$ Pauli spin matrices with
$\overline{\psi}_j = \hat\gamma_1\hat\psi_j^\dagger$ and
$\hat\gamma_++\hat\gamma_- = \hat\gamma_1$ providing the mixing
between the two chains.  As discussed in Ref.~\cite{Bittner:094909},
we used the highest occupied and lowest unoccupied ($\pi$ and $\pi^*$)
orbitals localized on each base as a orthonormal basis.  For the
single particle terms (representing electron and hole transfer between
bases), we use values reported by Anatram and Mehrez as determined by
computing the Coulomb integrals between HOMO and LUMO levels on
adjacent base pairs with in a double-strand B DNA sequence using
density functional theory (B3LYP/6-31G)~\cite{Mehrez:2005} taking the
geometries of each base from the B-DNA structure.  When $r_j^+ =
r_j^-$, Eq.~\ref{bandop} is identical to the Hamiltonian used by
Creutz and Horvath~\cite{Creutz:1994} to describe chiral symmetry in
quantum chromodynamics in which the terms proportional to $r$ are
introduced to make the ``doublers'' at $q\propto \pi$ heavier than the
states at $q \propto 0$.

Taking the chain to homogeneous and infinite in extent, one can easily
determine the energy spectrum of the valence and conduction bands by
diagonalizing
\begin{eqnarray}
\hat{h}_{1}=
\left(
\begin{array}{cc}
\epsilon_{a} + 2 t_{a} \cos(q)   &  h + r^{+}e^{-iq} + r^{-}e^{+iq} \\
h + r^{+}e^{+iq} + r^{-}e^{-iq} \    & \epsilon_{b} + 2t_{b} \cos(q)
\end{array}
\right)\label{bandop}
\end{eqnarray}
where $\epsilon_{a,b}$ and $t_{a,b}$ are the valence band or
conduction band site energies and intra-strand hopping integrals.
When $r_j^+ = r_j^-$, Eq.~\ref{bandop} is identical to the Hamiltonian
used by Creutz and Horvath~\cite{Creutz:1994} to describe chiral
symmetry in quantum chromodynamics in which the terms proportional to
$r$ are introduced to make the ``doublers'' at $q\propto \pi$ heavier
than the states at $q \propto 0$.  In particular, we note that when $t
= h/2r$, the band closes at $q = \pm\pi$ but has a gap at $q = 0$.

\begin{figure}
\begin{center}
\includegraphics[width=0.6\columnwidth]{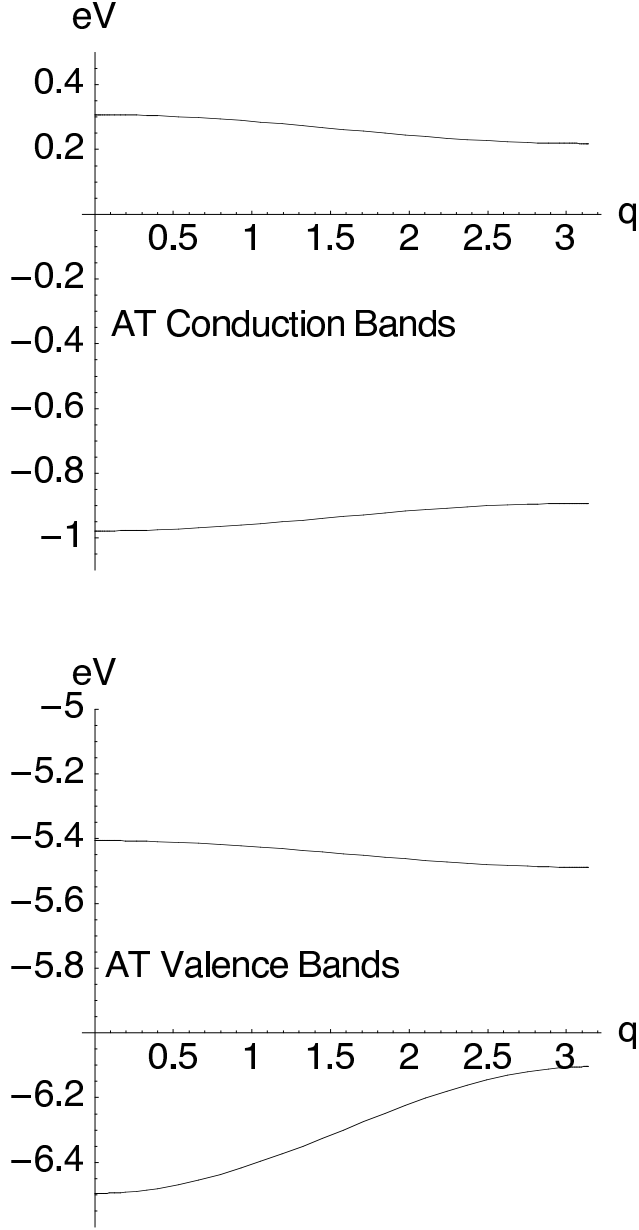}
\end{center}
\caption{(L) Band structure for homogeneous AT DNA chain.
 }\label{bands}
\end{figure}

The single particle parameters are taken from Anatram and Mehrez as
determined by computing the Coulomb integrals between HOMO and LUMO
levels on adjacent base pairs with in a double-strand B DNA sequence
using density functional theory (B3LYP/6-31G)~\cite{Mehrez:2005}.
%
Parameters used in our model are presented in Table ~\ref{params}.
 It is important to note thatg
 the asymmetry introduced with $r_j^+\ne r_j^-$ gives directionality
 between the 3'- and 5'- ends of the chain.  Introducing these
 parameters into Eq.~\ref{bandop} leads to 4 separate cosine shaped
 bands corresponding to conduction and valence bands localized along
 each chain as show in Fig.~\ref{bands}.


The coupling between the conduction and valence bands is accomplished
by introducing short-ranged Coulomb and exchange interactions as well
as ``dipole-dipole'' terms which couple geminate electron-hole pairs
on different sites and considering only single excitations,
\begin{eqnarray}
H(12) = h_1 + h_2 + \sum_{{\bf m,n}} V_{{\bf m,n}} A_{\bf m}^\dagger A_{\bf n}
\end{eqnarray}
where the $A_{\bf m}$ are spin-symmetrized composite operators that
create or remove singlet or triplet electron/hole pairs in
configuration $|{\bf m}\rangle = |i\overline{j}\rangle$ where $V_{\bf
mn} = -\langle m\overline{n}||n\overline{m}\rangle +
2\delta_{S0}\langle m \overline{n} || \overline{n}m\rangle$ where $S =
1,0$ is the total spin.
\cite{karabunarliev:4291,karabunarliev:3988,karabunarliev:10219,karabunarliev:057402}
In our model, we include three types of electron/hole interactions, an
on-site direct Coulomb $J = \langle
n\overline{n}||n\overline{n}\rangle $, an on-site exchange term
$K=\langle n\overline{n} |\overline{n}n\rangle$, and an inter-site
singlet exciton transfer term: $d_{mn} = \delta_{S0}\langle
m\overline{m}||n\overline{n}\rangle$.
 
 For simplicity, we take the on-site Coulomb interaction as $J =
\langle n\overline{n}||n\overline{n}\rangle $ and the on-site exchange
interaction $K=\langle n\overline{n} |\overline{n}n\rangle$ to be
adjustable parameters with values $J = 2.5$ eV and $K = 1.0$ eV for
both purines and pyrimidines.  We assume these interactions to be
local since the distance at which the Coulomb energy between an
electron/hole pair equals the thermal energy in aqueous ionic media at
300K is on the order of the base-stacking distance.  Furthermore, such
values are certainly in the correct range for conjugated cyclic
organic systems.
\cite{karabunarliev:4291,karabunarliev:3988,karabunarliev:10219,karabunarliev:057402}
Lastly, we estimated the coupling between geminate electron/hole pairs
on different bases $\langle n\overline{n} || m\overline{m}\rangle$ via
a point-dipole approximation by mapping the $\pi-\pi^*$ transition
moments onto the corresponding base in the B DNA chain.  These are
obtained from the isolated bases by performing single configuration
interaction (CIS) calculations using the GAMESS\cite{GAMESS} quantum
chemistry package on the corresponding 9-methylated purines and
1-methylated pyrimidines after optimizing the geometry at the
HF/6-31(d)G level of theory.

The use of the point-dipole approximation in this case is justified
mostly for convenience and given the close proximity of the bases,
multipole terms should be included in a more complete model.
\cite{Bouvier2002vu} As a result, the matrix elements used herein
provide an upper limit (in magnitude) of the couplings between
geminate electron-hole pairs.  Most importantly, however, the
point-dipole approximation provides a robust means of incorporating
the geometric arrangement of the bases into our
model.\cite{Bittner:094909}

 \begin{table}[h]
   \caption{Charge-transfer and exciton transfer terms for AT
   DNA.   (B-FORM) \label{params}}
 \begin{tabular}{lccc}
\hline
\hline
Description     & Value & Reference \\
\hline
Intrachain $e$ transfer: $A_i-A_{i+1}$  & $t_{ea} = +0.024$eV  &  Ref.~\cite{Mehrez:2005}\\ 
Intrachain $e$ transfer:  $T_i-T_{i+1}$  & $t_{eb} =  -0.023 $eV & Ref.~\cite{Mehrez:2005}\\ 
Interchain $e$ transfer:  $A_i-T_i$       & $h_e=  +0.063$ eV   & Ref.~\cite{Mehrez:2005}\\
Interchain 3'-3' $e$ transfer:  $A_i-T_{i+1} $&  $r_i^+ =  -0.012$ eV  & Ref.~\cite{Mehrez:2005}\\
Interchain 5'-5' $e$ transfer:  $A_{i}-T_{i-1} $&  $r_i^- = -0.016$ eV&  Ref.~\cite{Mehrez:2005}\\
\hline
Intrachain $h$ transfer:  $T_i-T_{i+1}$  & $t_{hb}=  -0.098$eV & Ref.~\cite{Mehrez:2005}\\ 
Intrachain $h$ transfer:  $A_i-A_{i+1}$  & $t_{ha} =  +0.021$eV&  Ref.~\cite{Mehrez:2005} \\ 
Interchain $h$ transfer:  $A_i-T_i$       & $h_h = +0.002$eV&  Ref.~\cite{Mehrez:2005}\\
Interchain 3'-3' $h$ transfer:  $A_i-T_{i+1} $&  $r_i^+ =  -0.007$ eV & Ref.~\cite{Mehrez:2005}\\
Interchain 5'-5' $h$ transfer:  $A_{i}-T_{i-1} $&  $r_i^- = +0.050$ eV & Ref.~\cite{Mehrez:2005}\\
\hline
Intrachain parallel dipole-dipole:  $T_i-T_{i+1}$  & $d^T_\parallel = 0.143$eV & Ref.~\cite{Bittner:094909}\\ 
Intrachain parallel dipole-dipole:  $A_i-A_{i+1}$  & $d^A_\parallel =0.0698$eV & Ref.~\cite{Bittner:094909}\\ 
Interchain perpendicular dipole-dipole:  $A_i-T_i$       & $d_\perp = -0.099$ & Ref.~\cite{Bittner:094909}\\
Interchain 3'-3' dipole-dipole:  $A_i-T_{i+1} $&  $d^+ = -0.013$eV& Ref.~\cite{Bittner:094909} \\
Interchain 5'-5' dipole-dipole:   $A_{i}-T_{i-1} $&  $d^- = -0.006$eV & Ref.~\cite{Bittner:094909}\\
\hline
 Site energies    & \\
 A(LUMO) & $\epsilon_{Ae} =  0.259 $eV   &Ref.~\cite{Mehrez:2005}\\
A(HOMO)& $\epsilon_{Ah} =  -5.45 $eV & Ref.~\cite{Mehrez:2005}\\
T(LUMO) &$\epsilon_{Te} =   -0.931 $eV &  Ref.~\cite{Mehrez:2005}\\
T(HOMO)&  $\epsilon_{Th} =  -6.298 $eV   &Ref.~\cite{Mehrez:2005} \\
 \end{tabular}
 \hrule
 \hrule

 \end{table}

\section{Excited State Dynamics: Homogeneous Lattice}
We now consider the fate of an initial singlet electron/hole pair
placed either in the middle of the thymidine side of the chain or the
adenosine side of the chain (i.e. a localized exciton with the
electron and hole starting on the same site).  We assume that such a
configuration is the result of an photoexcitation at the appropriate
photon energy (4.87 eV for the thymidine exciton and 5.21 eV for the
adenosine exciton respecitively) and based upon the observation that
the UV absorption spectra largely represents the weighted sum of the
UV spectra of the constituent bases.  Since these are not stationary
states, they evolve according to the time-dependent Schr{\"o}dinger
equation, which we integrate using the Tchebychev expansion of the
time-evolution operator.~\cite{tal-ezer:3967}

\begin{figure}[t]
\includegraphics[width=0.48\columnwidth]{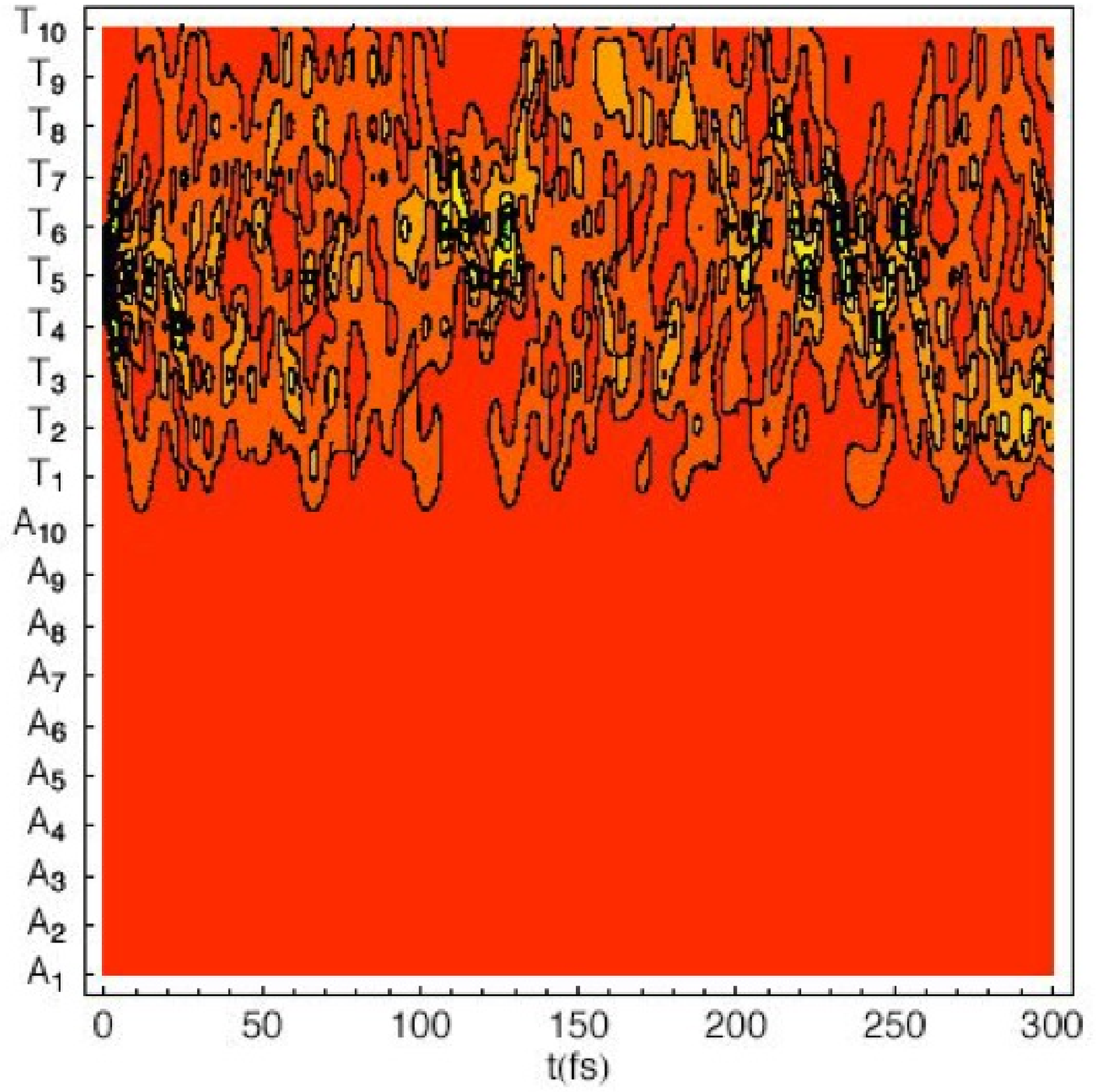}
\includegraphics[width=0.48\columnwidth]{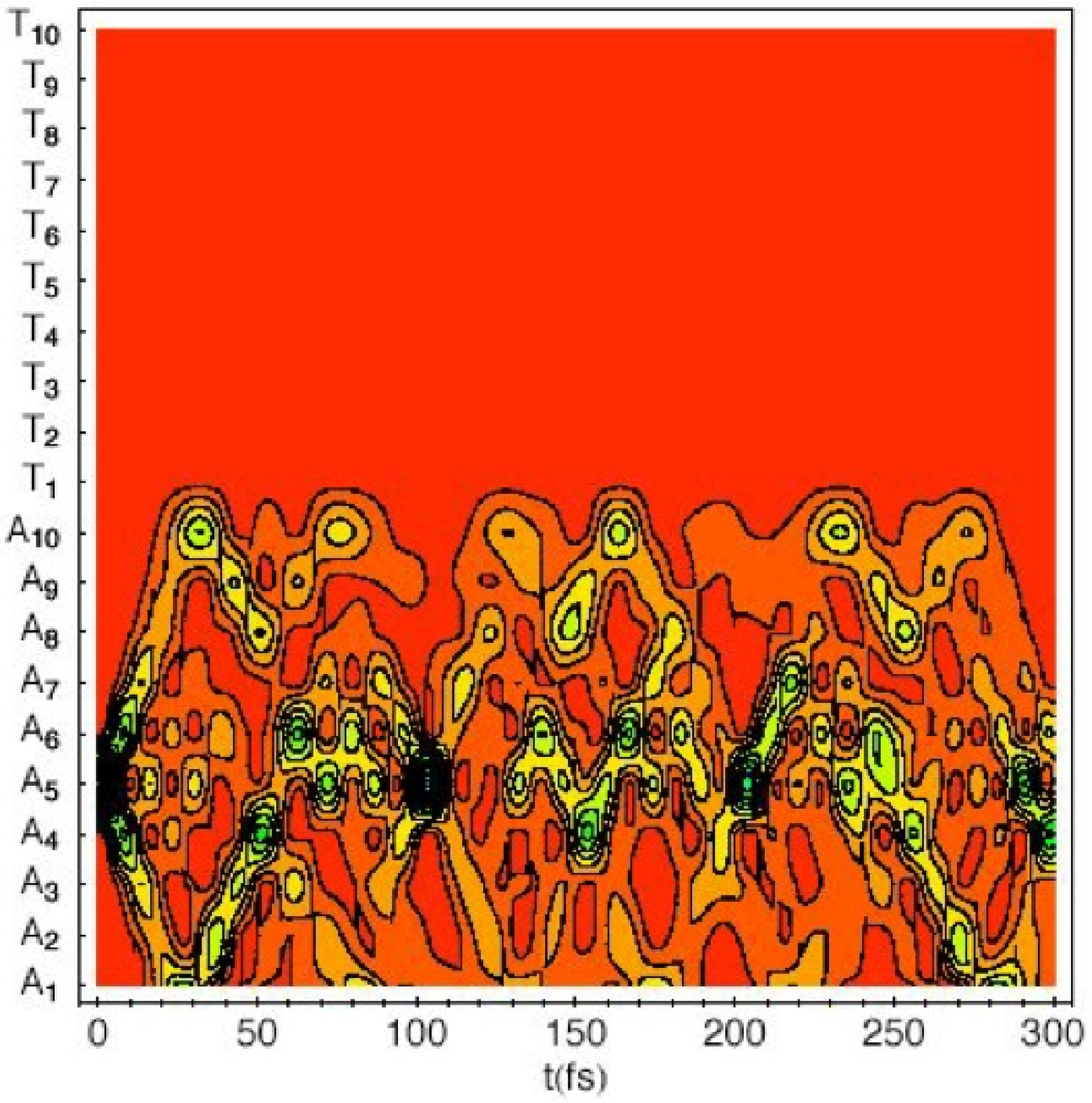}
\includegraphics[width=0.48\columnwidth]{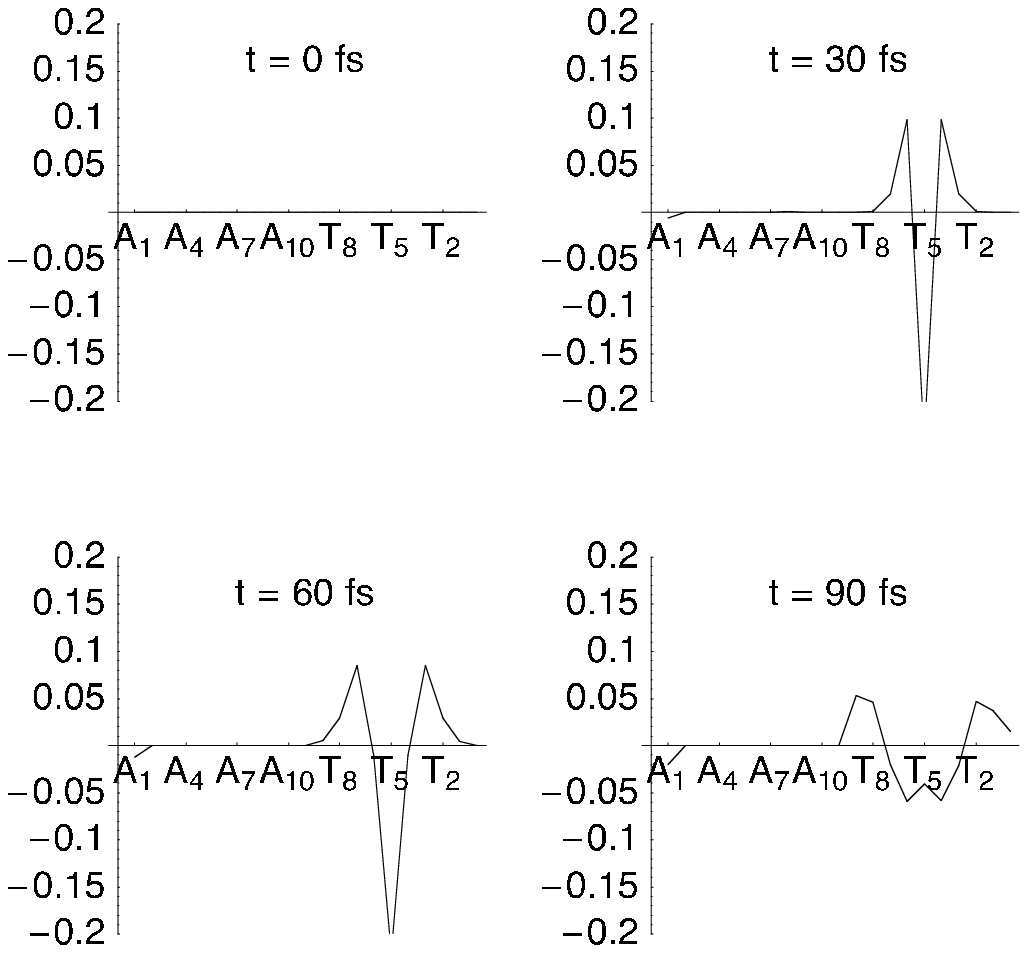}
\includegraphics[width=0.48\columnwidth]{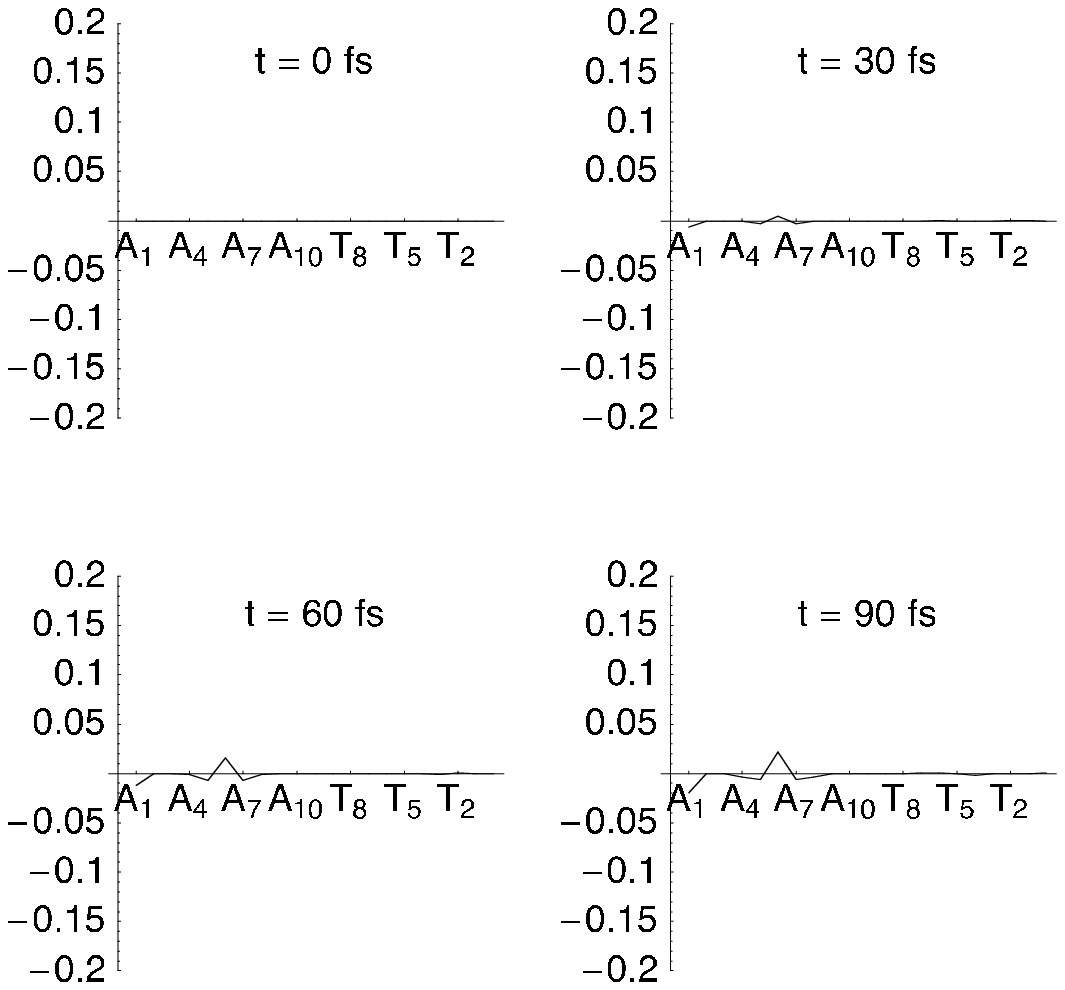}
\caption{Time evolution of excitonic density following excitation 
placed on $T_5$ (top-left) or $A_5$ (top-right) site.  The 
bottom panels show the net 
charge on each site following excitation of $T_5$ (bottom-left)  or 
$A_5$ (bottom-right) }
\label{fig2}
\end{figure}

In the top two frames of Fig.~\ref{fig2} we show the transient
probability for finding an exciton placed on the adenosine (left) or
thymidine (right) chain at time $t=0$ in some other excitonic
configuration along either the adenosine or thymidine chain at some
time $t$ later.  In both cases, negligible exciton density is
transferred between chains and the excitons rapidly become delocalized
and scatter ballistically down the DNA chain.

There are some striking differences, however, between the exciton
dynamics in adenosine versus those in thymidine.  First, in comparing
the $d_\parallel$ matrix elements, one easily concludes that the
exciton mobility along the thymidine chain is considerably greater
than the mobility along the adenosine chain.  This can be see in Fig
2. comparing the time required for an excitonic wavepacket to reach
the end of either chain.  In adenosine, the exciton travels nearly 5 base
pairs in about 25 fs where as an exciton along the thymidine chain
covers the same distance in about 10 fs.  This factor of two
difference in the exciton velocity is commensurate with the
$\approx$1:2 ratio of the $d_\perp^A:d_\perp^T$ intrachain excitonic
couplings.

Secondly, we note that the adenosine exciton remains qualitatively
more ``cohesive'' than the thymidine exciton showing a number of
ballistic traverses up and down the adenosine chain over the 300 fs we
performed the calculation.  One can also note that the exciton
velocity in the 5'-3' direction is slightly greater than in the 3'-5'
direction as evidenced by the exciton rebounds off site $A_1$ slightly
sooner than it rebounds from site $A_{10}$.  This is due to the
asymmetry introduced in by the $r^\pm$ and $d^\pm$ terms.  All in all,
one can clearly note a series of strong recurrences for finding the
adenosine exciton on the original site $A_5$ every 100 fs.  The
thymidine exciton dynamics are far more complex as the exciton rapidly
breaks apart.  While few recursions can be noted, however, after the
first ballistic traverse, the thymidine exciton no longer exists as a
cohesive wavepacket and is more or less uniformly distributed along
the thymidine side of the chain.

The excitonic dynamics only tell part of story.  In the lower 
frames of Fig. \ref{fig2} we show the net charge taken as the
difference between the hole density and electron density on a given
base.    In the case where the initial exciton is on the
adenosine chain, very little charge-separation occurs over the time
scale of our calculation.  On the other hand, when the exciton is
placed on the thymidine chain, the exciton almost {\em immediately}
evolves into a linear combination of excitonic and charge-separated
configurations.  What is also striking is that in neither case do
either the electron or hole transfer over to the other chain even
though energetically charge-separated states with the electron on the
thymidine and the hole on the adenosine sides of the chain are the
lowest energy states of our model.~\cite{Bittner:094909} It is
possible, that by including dissipation or decoherence into our
dynamics, such relaxation will occur, however, on a time scale
dictated by cross-chain transfer terms.  For the coupling terms at
hand, electron or hole transfer across base pairs occurs on the time
scale of 3-4 ps.

\begin{figure}[t]
\includegraphics[width=0.48\columnwidth]{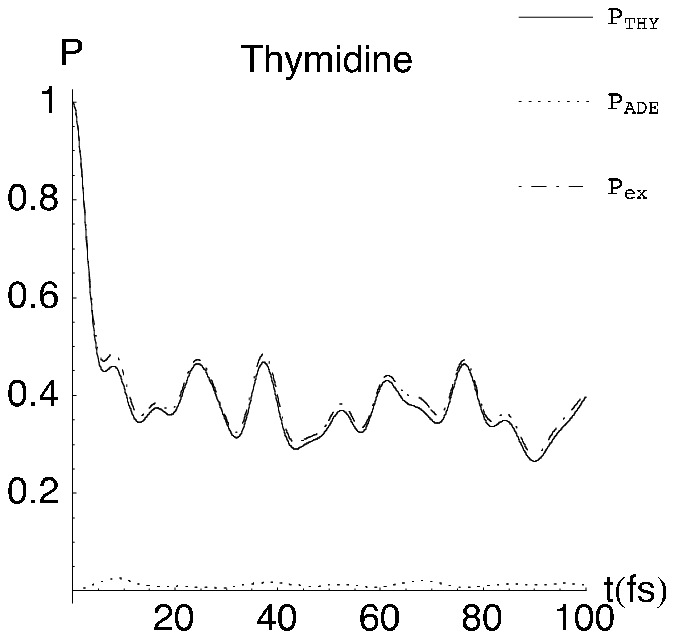}
\includegraphics[width=0.48\columnwidth]{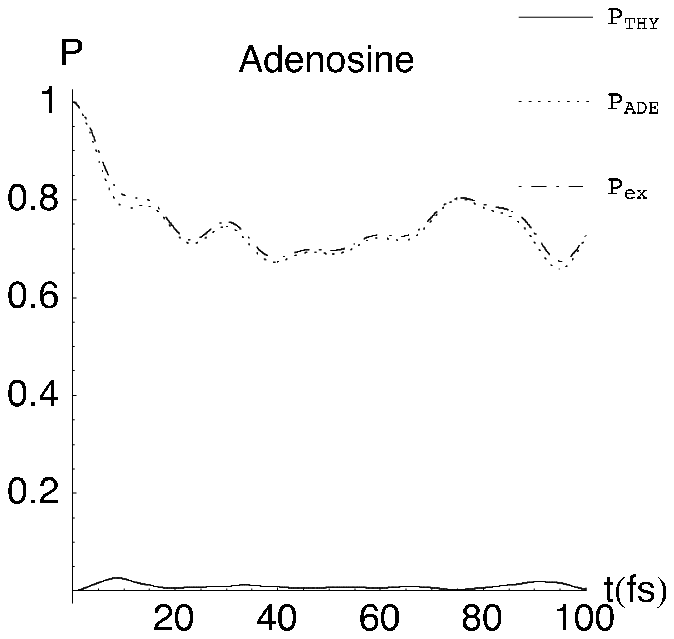}
\caption{Net probability for an exciton originating on the thymidine (left) or
adenosine (right) side of the chain to remain as an exciton on the original chain, 
to be found as an exciton on the other chain, and total exciton population.}
\label{fig3}
\end{figure}

The difference between the excitonic dynamics following excitation of
A vs. T can be quantitatively noted by comparing the curves shown in
Fig.~\ref{fig3} where we compare the projection of the time-evolved
state onto the excitonic configurations of the chain on which the
exciton was placed ($P_{AA}$ and $P_{TT}$) compared to the projection
onto the excitonic configurations of the other chain ($P_{AT}$ and
$P_{TA}$).  For the case in which the adenosine chain was excited,
approximately 75\% of the total probability density remains as
excitonic configurations along the adenosine chain.  In stark
contrast, only about 40\% of the initial thymidine exciton remains
excitonic along the thymidine side of the chain.  The reason for the
remarkable difference between the two chains stems from the difference
in electron and hole mobility along the thymidine chain.  Indeed,
comparing the electron and hole hopping terms given above,
$t_{h}/t_{e} \approx 4$ for the thymidine chain compared to
$t_{h}/t_{e} \approx 1$ for along the adenosine chain.  This is
manifest in the lower left panel of Fig.~\ref{fig2} where we see almost
immediately a negative charge remaining for a few fs on the site where
the initial excitation was placed.   In contrast, no charge-separation 
is seen following excitation along the adenosine side of the chain.

\section{Excited State Dynamics: Effects of Static Disorder}

A proper model of the electronic dynamics in DNA must include some
contribution from the solvent and environment.  For DNA in water at
300K, we assume that the electronic processes described in our model
are fast compared to the structural and environmental fluctuations the
DNA lattice itself such that the parameters for the B-DNA structure
represent the average values for the system in aqueous media.  For
example, the characteristic time scale for the relative lateral and
longitudinal motions of bases in DNA is 10 to 100 fs with amplitudes
of 0.01 to 0.1 nm. ~\cite{McCammon:1987} Since electronic interactions
between bases are sensitive to the fluctuations in the geometry of the
DNA structure.  factors such as salt concentration and other solvent
media will have profound impact on the structure and hence on the
model parameters.

Since short strands of DNA are fairly rigid, the electronic coupling
terms are likely most sensitive to the upon the dihedral angle,
$\theta_{ij}$ between adjacent bases.  If we take the fluctuations in
$\theta_{ij}$ to be $\delta\theta^2 = k_BT/I\Omega^2$ where $I$ is the
reduced moment of inertia of the AT base-pair and $\Omega = 25cm^{-1}$
is the torsional frequency.  This gives an RMS angular fluctuation of
about 5\% about the avg. $\overline{\theta_{i,i+1}}=35.4^\circ$
helical angle.  Since this is a small angular deviation, we take the
fluctuations in the electronic terms to be proportional to
$\delta\theta^2$ and sample these terms from normal distributions
about B-DNA average values.

\begin{figure}[t]
\includegraphics[width=0.48\columnwidth]{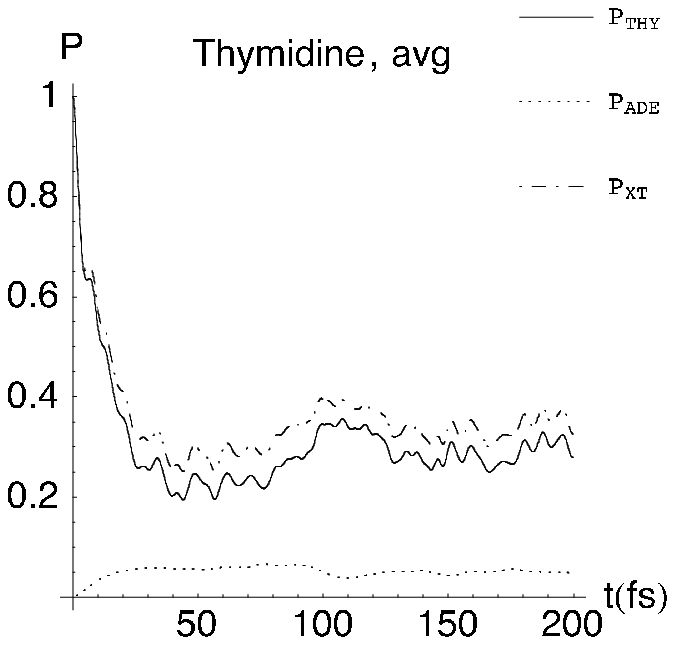}
\includegraphics[width=0.48\columnwidth]{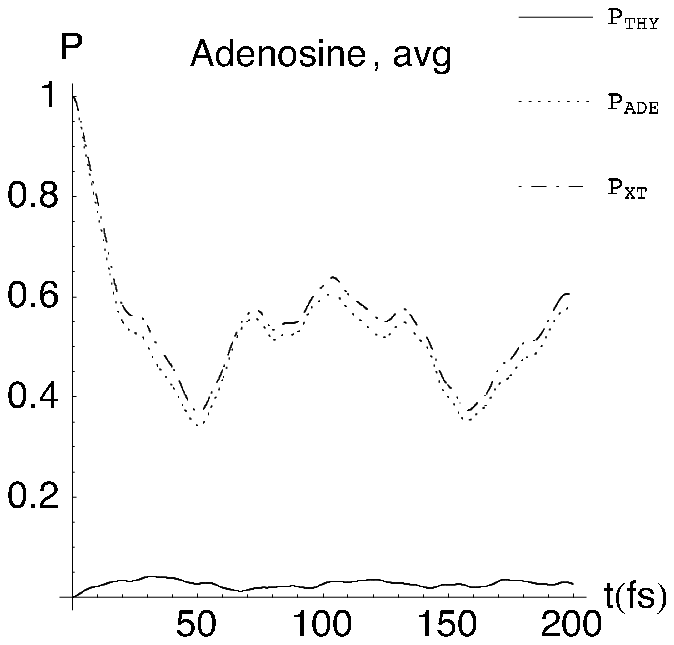}
\caption{Same as in Fig.~\ref{fig3} except averaging over ensemble of 
configurations.}
\label{fig4}
\end{figure}

The effect of disorder on the excitonic dynamics are seen in Fig.~\ref{fig4}.
As in case of the homogeneous lattice above, our initial state is a
Frenkel exciton on thymidine \#5. The right-hand plot of Fig.~\ref{fig4} shows
the probability of finding the time-evolved state in various excitonic
configurations.  While the time-scale for the exciton break up is
slower than in the homogenous lattice, one can deduce that even with
static off-diagonal disorder, the thymidine exciton is subject to
dissociate to charge-transfer (i.e. polaron) pairs on an ultrafast
timescale.   The adenosine exciton appears to be more profoundly 
affected by the lattice disorder with some population transfered to the 
thymidine chain.  Since the total exciton population is lowered upon 
introducing disorder, the remaining excited state population exists as
charge-separated pairs.   Analysis of the transient excited state wavepackets (not-shown)
indicates a small but significant fraction of  $A_i^+-T_i^-$ interstrand
charge-transfer configurations.

\begin{figure}[t]
\includegraphics[width=0.48\columnwidth]{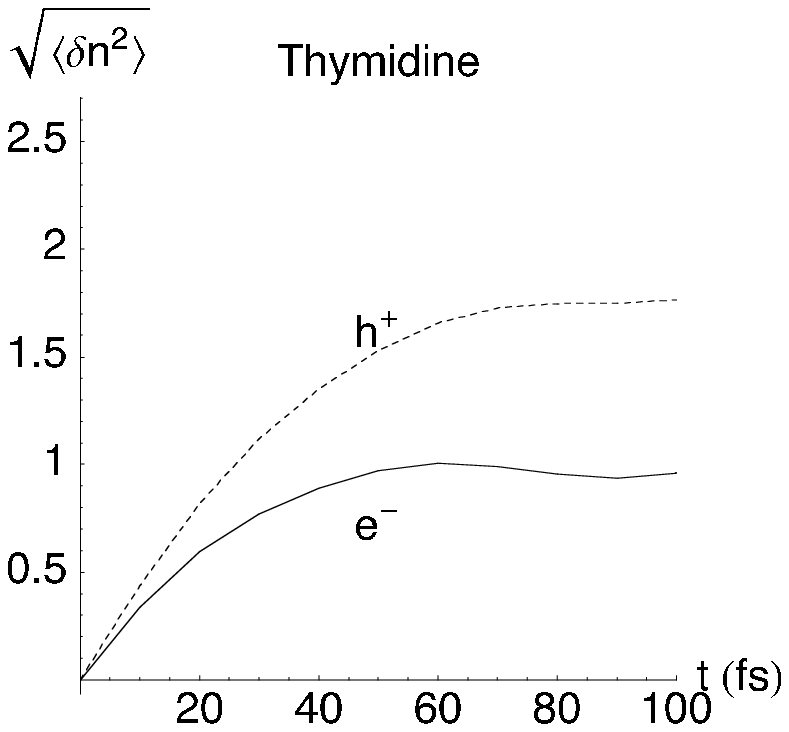}
\includegraphics[width=0.48\columnwidth]{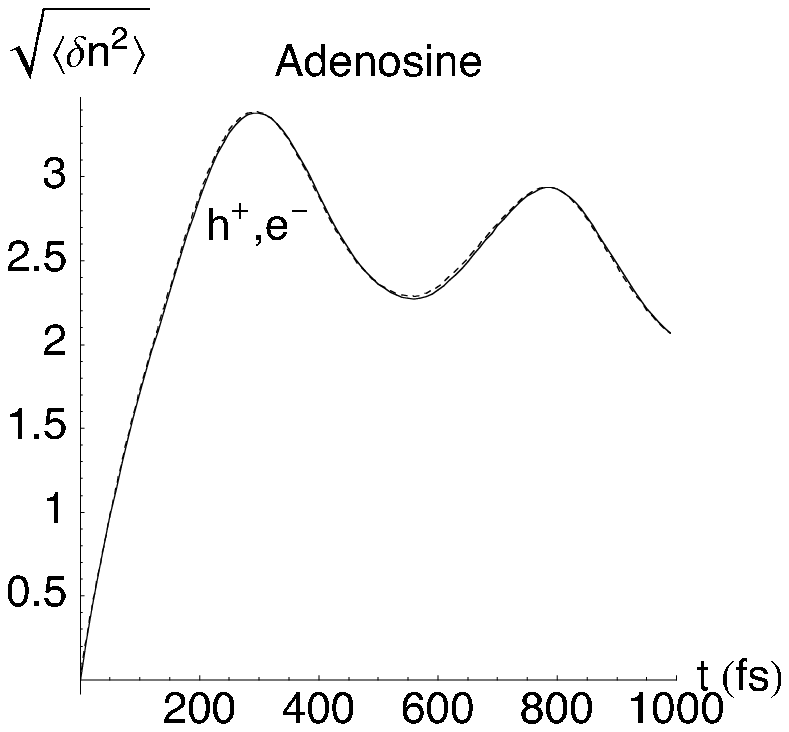}
\includegraphics[width=0.48\columnwidth]{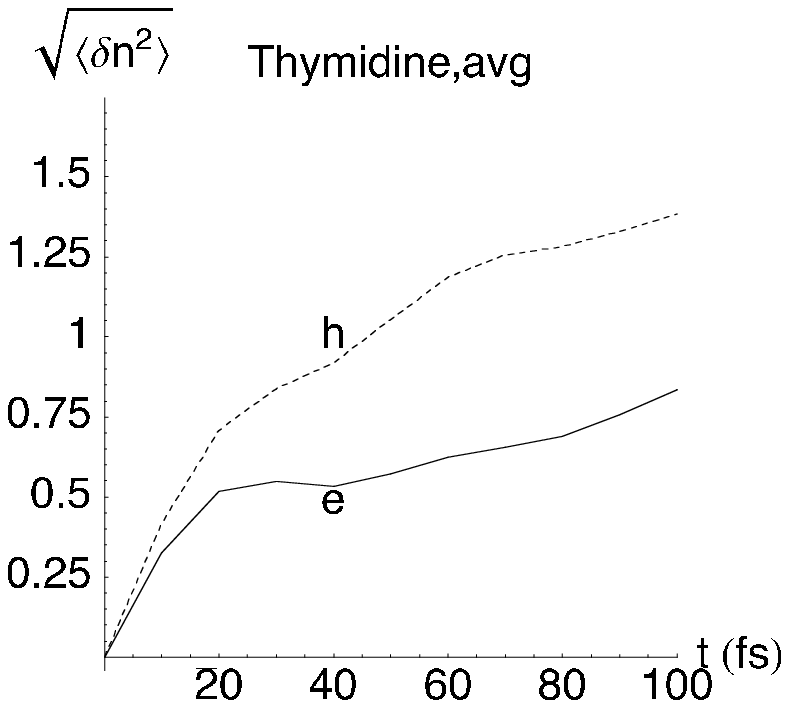}
\includegraphics[width=0.48\columnwidth]{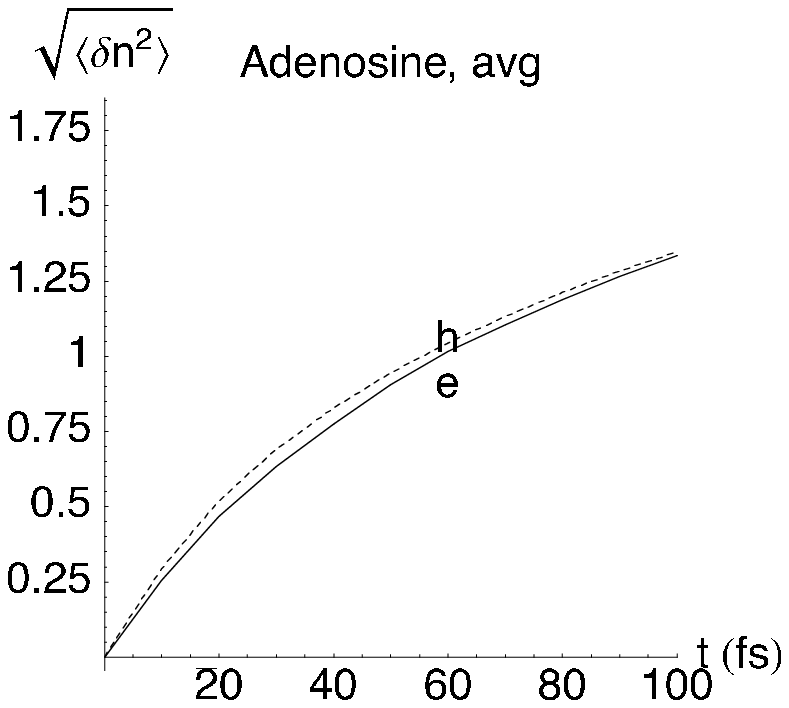}
\caption{RMS width of electron and hole densities vs. time for 
excitations originating on different sides of the AT chain.  
Top: Homogeneous chain, Bottom: averaging over
20 random lattice configurations.}\label{fig6}
\end{figure}

Lastly, we examine the relative mobilities of the electron and hole by calculating the 
r.m.s. width of the electron and hole densities on either chain following excitation, 
the results of which are shown in Fig.~\ref{fig6}.  In the top two plots are the 
r.m.s. widths for the homogeneous lattice following excitation.  As discussed above, the 
hole is considerably more mobile than the electron along the thymidine side of the chain.  
This is very much evidenced by the more rapid spread of the hole density compared to the 
the electron density.  Likewise,  along the adenosine side, the conduction and valence bands
have nearly identical band-widths and the electron  and hole densities evolve very much 
in concert.     This scenario is preserved upon even averaging over configurations indicating 
that to within a 5\% error in our parameter set or the to within the typical thermal 
fluctuations of a DNA chain, our predicted dynamics are quite robust.

\section{Summary}

The results described herein paint a similar picture to that described
by recent ultrafast spectroscopic investigations of (dA).(dT)
oligomers in that the initial excitonic dynamics is dominated by
base-stacking type interactions rather than by inter-base couplings.
Interchain transfer is multiple orders of magnitude slower than the
intrachain transport of both geminate electron/hole pairs as excitons
and independent charge-separated species.  Indeed, for an exciton
placed on the adenosine chain, our model predicts that exciton remains
as a largely cohesive and geminate electron/hole pair wave function as
it scatters along the adenosine side of the chain.  Our model also
highlights how the difference between the mobilities in the conduction
and valence bands localized along each chain impact the excitonic
dynamics by facilitating the break up of the thymidine exciton into
separate mobile charge-carriers.  In the actual physical system, the
mobility of the free electron and hole along the chain will certainly
be dressed by the polarization of the medium and reorganization of the
lattice such that the coherent transport depicted here will be
replaced by incoherent hopping between bases.

The significance of the breakup of the exciton is twofold.  First, it
 is well recognized that photoexcitation of adjacent stacked
 pyrimidine bases leads to the formation of {\em cis-syn} cyclobutane
 pyrimidine-dimer lesions.  However, this dimerization occurs only in
 the triplet (rather than singlet) excited state. \cite{Sinha:2002}
 Consequently, spin-flip must occur {\em ether} via spin-orbit
 coupling or via recombination of polaron
 pairs.\cite{karabunarliev:057402} If we assume that the spins are
 decorrelated at some intermediate distance $r\propto e^2/\epsilon kT$
 where the Coulomb energy is equal to the thermal energy,
 photoexcitation of a thymine sequence could rapidly result in a
 population of triplet excitons formed by exciton dissociation
 followed by geminate recombination.  Secondly, the process is
 reversible and triplet reactivation of the dimer can lead to repair
 of the lesion.

Isolating the photoexcitation to the originally excited chain
minimizes the potential mutagenenic damage to the DNA sequence since
it preserves the complementary chain as an undamaged back-up copy of
the genetic information.  It is fascinating to speculate whether or
not the isolation of a photoexcitation and its photoproducts to the
original chain was an early evolutionary selection criteria for the
eventual emergence of DNA as the carrier of genetic information.

In conclusion, we present herein a rather compelling model for the
short-time dynamics of the excited states in DNA chains that
incorporates both charge-transfer and excitonic transfer.  It is
certainly not a complete model and parametric refinements are
warranted before quantitative predictions can be established.  For
certain, there are various potentially important contributions we have
left out: disorder in the system, the fluctuations and vibrations of
the lattice, polarization of the media, dissipation, quantum
decoherence.  We hope that this work serves as a starting point for
including these physical interactions into a more comprehensive
description of this system.

\section{Acknowledgments} 
This work was funded in part by grants from the Robert Welch
Foundation, and the National Science Foundation (CHE-0345324 ).


\end{document}